\documentclass[twocolumn,pra,amsmath,amssymb,unsortedaddress,showpacs,nofootinbib]{revtex4}

\usepackage{latexsym,epsfig,graphicx}
\usepackage{dcolumn}
\usepackage{bm}
\usepackage{color} 
\usepackage{graphicx}
\usepackage{subfigure}

\begin{document}

\title{Pair tunneling, phase separation and dimensional crossover in imbalanced fermionic superfluids
in a coupled array of tubes}

\author{Kuei Sun}
\author{C. J. Bolech}\affiliation {Department of Physics, University of Cincinnati, Cincinnati,
Ohio 45221-0011, USA}
\date{May 29, 2013}


\begin{abstract}
We study imbalanced fermionic superfluids in an array of
one-dimensional tubes at the incipient dimensional crossover
regime, wherein particles can tunnel between neighboring tubes. In
addition to single-particle tunneling (ST), we consider pair
tunneling (PT) that incorporates the interaction effect during the
tunneling process. We find that with an increase of PT strength, a
system of low global polarization evolves from a structure with a
central Fulde-Ferrell-Larkin-Ovchinnikov (FFLO) state to one with
a central BCS-like fully-paired state. For the case of high global
polarization, the central region exhibits pairing zeros embedded
in a fully paired order. In both cases, PT enhances the pairing
gap, suppresses the FFLO order, and leads to spatial separation of
fully paired and fully polarized regions, the same as in higher
dimensions. Thus, we show that PT beyond second-order ST processes
is of relevance to the development of signatures characteristic of
the incipience of the dimensional crossover.
\end{abstract}
\maketitle

\begin{figure}[t]
\centering
\includegraphics[width=6.5cm]{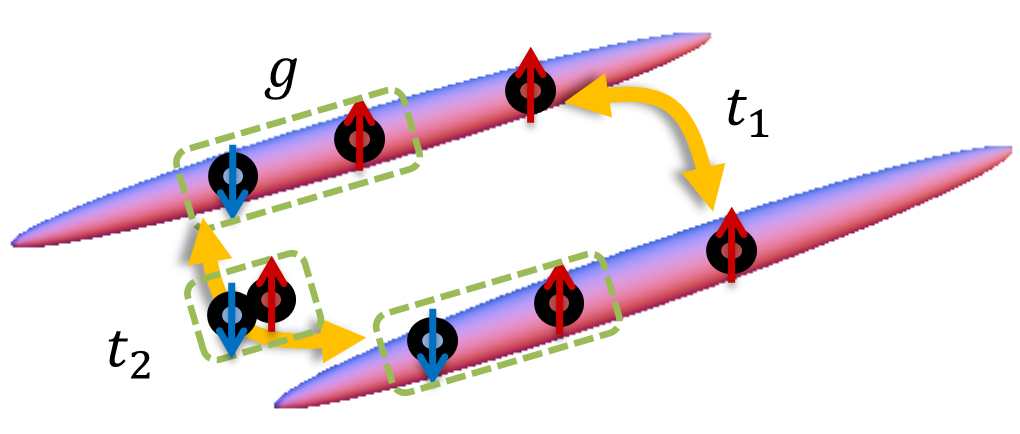}
  \caption{(Color online) Illustration of single particle
and pair tunneling processes $t_1$ and $t_2$, respectively,
between two neighboring tubes in the array. On each tube, two
opposite-spin atoms can form a bound pair (circled in the graph)
in the presence of an attractive interaction $g$. The circled pair
in the middle indicates that the two atoms remain paired during
the $t_2$ process.}
        \label{fig:f01}
\end{figure}

\section{Introduction}\label{sec:introduction}
Superconductivity and ferromagnetism are two ubiquitous but
competing phenomena in condensed matter systems. Spin imbalance
and magnetic fields induced by ferromagnetism tend to suppress
Cooper pairing, which is responsible for superconductivity. For
more than four decades, an interesting phase, the
Fulde-Ferrell-Larkin-Ovchinnikov (FFLO) state~\cite{FF,LO}, has
been suggested as the concurrence of both ferromagnetic and
oscillatory superconducting orders~\cite{Casalbuoni04,Buzdin05},
but its direct confirmation is still elusive. Recently, owing to
the capability of controlling particle densities, tuning
interactions, and cooling into quantum
degeneracy~\cite{Bloch08,Ketterle08}, cold atomic systems have
become a promising platform for searching for FFLO
order~\cite{Giorgini08,Radzihovsky10}. In experiments of trapped
Fermi gases, density profiles that reflect the interplay of spin
imbalance (ferromagnetic order) and Cooper pairing have been
observed~\cite{Zwierlein06,Partridge06,Liao10}. In addition,
experiments have also revealed a significant dimensional
dependence of the profiles: in three dimensions (3D) a fully
paired profile takes place at the trap center and a polarized
profile does off center~\cite{Zwierlein06,Partridge06}, while in
1D the central region is always polarized~\cite{Liao10}. These
observations agree with theoretical studies of the FFLO state in
1D and the trap-induced phase separation in
3D~\cite{Mizushima05,Sheehy06,Orso07,Liu07,Feiguin07,Kakashvili09,Kim11,Baksmaty11},
but the marked difference between these two limits also raises the
need for understanding the intermediate regime. Several works have
focused on the dimensional crossover regime of various kinds of
continuous
systems~\cite{Yang01,Parish07,Zhao08,Devreese11,Lutchyn11,Sun12}
or Hubbard lattices~\cite{Feiguin09,Kim12}, but with a different
emphasis than the present work.

In this paper, we study a realizable system of a two-dimensional
optical lattice array of one-dimensional tubes, subject to a
global trapping potential~\cite{Moritz05,Liao10,Hulet12}. The
incipient dimensional crossover regime of this system, which can
be experimentally accessed by gradually lowering the lattice
depth, is modeled by incorporating the kinetics of single-particle
tunneling (ST) as well as a key ingredient representing the
tunneling of paired opposite-spin atoms---pair tunneling
(PT)---between neighboring tubes (as illustrated in
Fig.~\ref{fig:f01}). The ST leads to an interesting magnetic
compressible-incompressible phase transition analogous to that in
the Bose-Hubbard model (discussed in Ref.~\cite{Sun12}), but is
not responsible for certain observed signatures in the profiles at
the dimensional crossover regime (which will be shown below). By
considering PT, we are able to describe the incipient evolution of
profiles from 1D toward 3D and obtain the emerging signatures of
the dimensional crossover at various global polarizations, such as
the inversion of the fully paired and polarized centers as well as
the growing spatial separation between the fully paired and fully
polarized regions.

The paper is outlined as follows. In Sec.~\ref{sec:pair} we
discuss the microscopic physical cause of the PT and evaluate its
strength using a two-channel model. In Sec.~\ref{sec:BdG}, we
construct a model Hamiltonian and apply a Bogoliubov--de Gennes
(BdG) treatment to solve for the density and pairing profiles of
the system. In Sec.~\ref{sec:result} we present the results and
discuss their physical meanings associated with PT. Finally, we
summarize our work in Sec.~\ref{sec:conclusion}.

\begin{figure}[t]
\centering
\includegraphics[width=6.5cm]{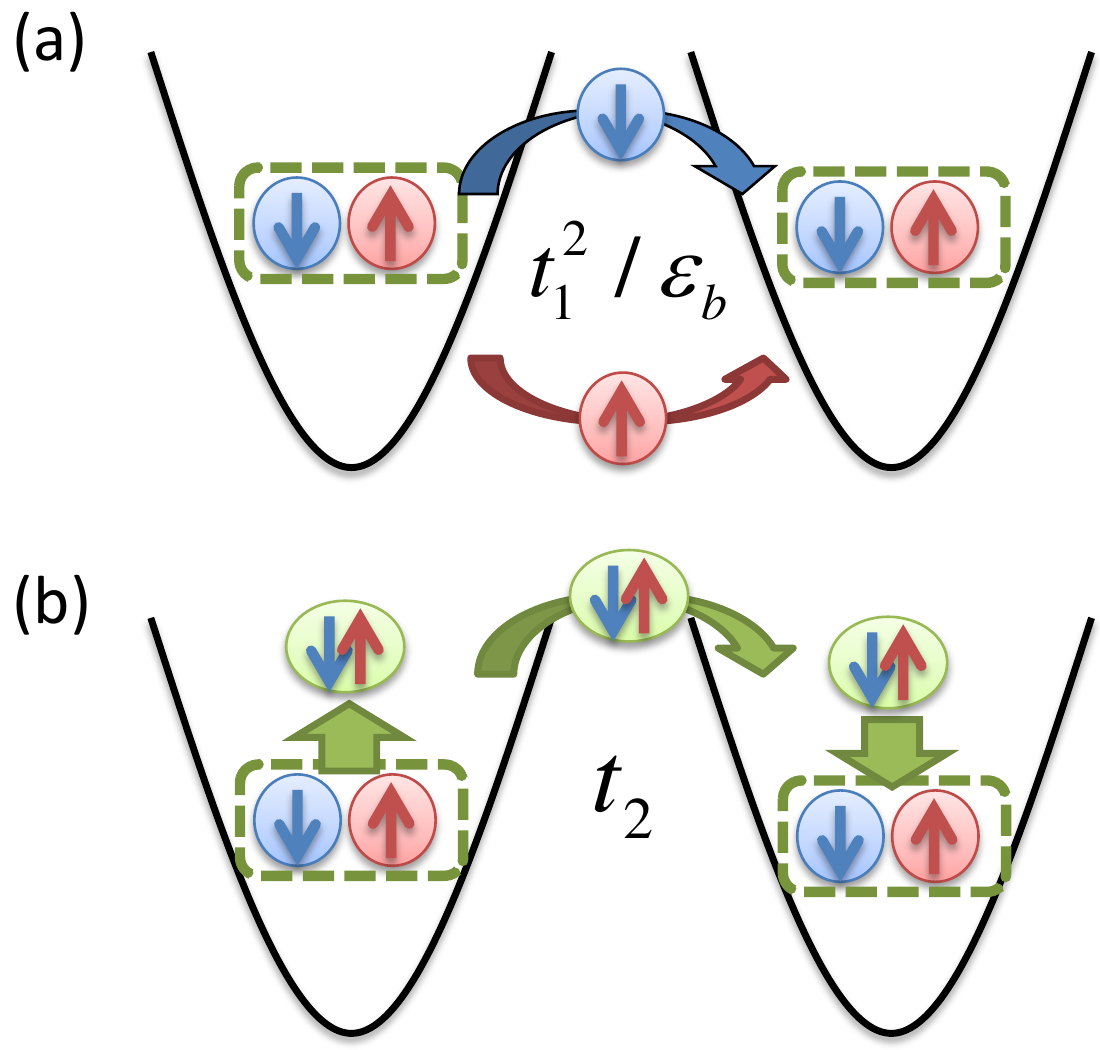}
  \caption{(Color online) Illustration of two microscopic kinetics
  of the pair tunneling in optical lattices. (a) Two atoms of
  a pair split, separately tunnel to the other site, and rebind,
  as a second-order process of strength $t_1^2/\epsilon_b$
  in the single-particle tunneling model.
  (b) Atoms bind via a pervasive Feshbach resonant effect
  during the whole tunneling event,
  with the effective pair tunneling strength $t_2$.}
        \label{fig:f02}
\end{figure}

\section{Physics of pair tunneling}\label{sec:pair}
For free fermions in a lattice potential, the intersite kinetics
is well described by ST processes of strength
$t_1$~\cite{Jaksch05}. For an attractively interacting case where
two opposite-spin atoms form a pair of binding energy
$\epsilon_b$, the kinetics of the pair tunneling would, in
principle, be incorporated as a process in which the two atoms of
a pair split, separately tunnel to the other site and rebind [see
Fig.~\ref{fig:f02}(a)]. This is contained in the ST model as a
second order process of strength $t_1^2/\epsilon_b$ and accounts
for the Josephson phenomena in the presence of superfluid orders,
cf.~\cite{Zhao08}.

However, in cold-atom experiments, the interaction is induced via
a Feshbach resonance~\cite{Chin10}, which is controlled by the
tuning of a magnetic field affecting the hyperfine energy
splittings. Because the field is applied throughout the system,
the interaction that leads to pairing exists on and \emph{between}
lattice sites. Therefore, we expect that atoms that remain paired
during the whole tunneling event can be another viable process
[see Fig.~\ref{fig:f02}(b)]. Such a process can be described as
tunneling of the paired atoms, with strength denoted as $t_2$.

One can estimate $t_2$ around the Feshbach resonant regime using a
two-channel model~\cite{Ketterle08,Giorgini08,Chin10} that
incorporates atomic and molecular degrees of freedom,
$\psi_\sigma$ and $\phi$, respectively. In optical
lattices~\cite{Dickerscheid05,Duan05,Wall12}, the partition
function of the system is
\begin{eqnarray}
\mathcal{Z} = \int {D\{ {\psi _{\sigma i}},{{\bar \psi }_{\sigma
i}}\} D\{ {\phi _i},{{\bar \phi }_i}\} {{\mathop{\rm e}\nolimits}
^{ - \int {d\tau ({S_a} + {S_m})} }}},\label{eqn:PAR1}
\end{eqnarray}
where ${S_a}$ contains terms associated only with the atomic
degrees of freedom, including the atomic tunneling as well as any
bare interatomic interaction, and
\begin{eqnarray}
{S_m} &=&  - {t_m}\sum\limits_{\left\langle {ij} \right\rangle}
{{{\bar \phi }_i}{\phi _j}}  - {\mu
_m}\sum\limits_{i} {{{\bar \phi }_i}{\phi _i}}\nonumber\\
&& + {U_{am}}\sum\limits_i {\left( {{{\bar \phi }_i}{\psi _{
\downarrow i}}{\psi _{ \uparrow i}} + {\rm{H}}{\rm{.c}}{\rm{.}}}
\right)}
\end{eqnarray}
involves the molecular tunneling $t_m$, molecular chemical
potential $\mu_m$, and atom-molecule coupling $U_{am}$. Here we
assume the intermolecular interaction is weak such that a
mean-field approximation ${{\bar \phi }_i}{{\bar \phi }_i}{\phi
_i}{\phi _i} \to \left\langle {{{\bar \phi }_i}{\phi _i}}
\right\rangle {{\bar \phi }_i}{\phi _i}$ can be applied to
incorporate the interaction as effective contributions to the
chemical potentials. We integrate out the molecular variable
$\phi$ in Eq.~(\ref{eqn:PAR1}) and obtain
\begin{eqnarray}
\mathcal{Z} = \int {D\{ {\psi _{\sigma i}},{{\bar \psi }_{\sigma
i}}\} {{\mathop{\rm e}\nolimits} ^{ - \int {d\tau ({S_a} +
{S'_a})} }}},
\end{eqnarray}
where $S'_a$ is expanded as
\begin{eqnarray}
{S'_a} &=& \frac{{U_{am}^2}}{{{\mu _m}}}\bigg[  - \sum\limits_i
{{{\bar \psi }_{ \uparrow i}}{{\bar \psi }_{ \downarrow i}}{\psi
_{ \downarrow i}}{\psi _{ \uparrow i}}}  \nonumber\\
&& - \frac{{{t_m}}}{{{\mu _m}}}\sum\limits_{\left\langle {ij}
\right\rangle} {{{\bar \psi }_{ \uparrow i}}{{\bar \psi }_{
\downarrow i}}{\psi _{ \downarrow j}}{\psi _{ \uparrow j}} +
\mathcal{O}\left( {\frac{{t_m^2}}{{\mu _m^2}}} \right)}
\bigg].\label{eqn:Sa2}
\end{eqnarray}
The first term in Eq.~(\ref{eqn:Sa2}) can be treated as a resonant
contribution to the inter-atomic interaction\footnote{In
combination with any bare interatomic interaction present, one
would obtain an effective interatomic interaction as discussed in
Ref.~\cite{Dickerscheid05}.}, while the second one appears as PT.
Therefore, we obtain
\begin{eqnarray}
{t_2} = \frac{{U_{am}^2}}{{\mu _m^2}}{t_m}. \label{eqn:t2}
\end{eqnarray}
By considering the tunneling strength in optical lattices given by
$\frac{4}{{\sqrt \pi }}{E_R}{\left( {{{{V_0}}}/{{{E_R}}}}
\right)^{3/4}}\exp [ - 2\sqrt {{{{V_0}}}/{{{E_R}}}} ]$ with $V_0$
the optical-lattice depth and $E_R$ the recoil
energy~\cite{Jaksch98,Buchler03}, we find
\begin{eqnarray}
\frac{{{t_2}}}{{{t_1}}} = \sqrt 2 \frac{{U_{am}^2}}{{\mu
_m^2}}\exp \left[ { - 2\sqrt {\frac{{{V_0}}}{{{E_R}}}} } \right]
\end{eqnarray}
This expression shows that $t_2$ has the same sign as $t_1$ and
can vary at fixed $t_1$ (or fixed lattice geometry) through the
tuning of $U_{am}$ and $\mu _m$. We also see that even if the
molecular tunneling is smaller than the atomic tunneling
($t_m<t_1$), $t_2$ can be comparable to or even larger than $t_1$
in relatively shallow lattices and near the Feshbach resonance
(large $U_{am}$ and small $\mu_m$~\cite{Bruun04}). Realistic
values can be estimated as ${U_{am}} = \sqrt {4\pi {\hbar
^2}\delta \mu W\left| {{a_{b}}} \right|/m} \int {{w_m}({\bf{r}} -
{{\bf{r}}_i})w_a^2({\bf{r}} - {{\bf{r}}_i})d{\bf{r}}}$ and ${\mu
_m} = \delta \mu \delta B$ (taking the bare molecular
limit)~\cite{Dickerscheid05,Duan05,Chin10}, where $\delta \mu$ is
the differential magnetic moment, $W$ is the resonance width,
$a_b$ is the background scattering length, $w_m$ ($w_a$) is the
molecular (atomic) Wannier wave function on site $i$, and $\delta
B$ is the detuning of the magnetic field. In an ongoing experiment
using a setup as in Ref.~\cite{Liao10} with $V_0 \sim 7
E_R$~\cite{Hulet12}, we can expect $t_2/t_1 \sim 1$ given
$|U_{am}/\mu_m| \sim 10$ (or $\delta B < 1.6$G). We remark that
perturbative renormalization-group analysis starting from 1D
becomes unreliable in this parameter regime.

In addition, our system shows pervasive pairing effects and is
thus different from an array of Josephson junctions that lack the
pairing mechanism in the insulating barriers between the
superconductors. Therefore, by providing extra channels for
Josephson-type tunneling, the PT processes can lead to an
enhancement of the superfluid order and its cross-tube coherence,
anticipating the dimensional crossover regime. Note that we focus
here on the leading processes in the lowest Bloch band, which has
been shown capable of describing well the realized optical lattice
systems~\cite{Jaksch98}. We point out that incorporating
higher-order, higher-band, or interband processes are possible
immediate extensions of our model~\cite{Duan05,Mathy09}, necessary
to explore even wider optical-lattice regimes.

\section{BdG calculation}\label{sec:BdG}
Incorporating both the ST and PT effects, the tube lattices
occupied by up-spin (majority) and down-spin (minority) atoms
${{\hat \psi _{\sigma=\uparrow/\downarrow,{\bf{r}}}(z)}}$ are
hence described by the microscopic Hamiltonian
\begin{eqnarray}
H &=& \int_z {\sum\limits_{\bf{r}} {\big( {\sum\limits_\sigma
{\hat \psi _{\sigma{\bf{r}}} ^\dag H_\sigma ^0{{\hat \psi
}_{\sigma{\bf{r}}} }}  - g\hat \psi _ {\uparrow{\bf{r}}} ^\dag
\hat \psi _ {\downarrow{\bf{r}}} ^\dag {{\hat \psi }_
{\downarrow{\bf{r}}} }{{\hat \psi }_ {\uparrow{\bf{r}}} }} \big)} }  \nonumber\\
&{+}&  \int_z {\sum\limits_{\left\langle {{\bf{rr'}}}
\right\rangle } {\big( {{-t_1}\sum\limits_\sigma  {\hat \psi
_{\sigma {\bf{r'}}}^\dag {{\hat \psi }_{\sigma {\bf{r}}}}}
{-t_2}\hat \psi _{ \uparrow {\bf{r'}}}^\dag \hat \psi _{
\downarrow {\bf{r'}}}^\dag {{\hat \psi }_{ \downarrow
{\bf{r}}}}{{\hat \psi }_{ \uparrow {\bf{r}}}}} \big)} },
\label{eqn:HAM}
\end{eqnarray}
with the $\hat z$ direction along the tube's axis and $
{\bf{r}}=(x,y)$ denoting tube indexes in the plane perpendicular
to $\hat z$. The one-particle Hamiltonian $H_\sigma ^0 = - ({\hbar
^2}/2m)\partial _z^2 + m(\omega _r^2{r^2} + \omega {z^2})/2 - {\mu
_\sigma }$ includes the kinetic energy in the $\hat z$ direction,
the global trapping potential, and the spin-dependent chemical
potentials. The on-tube coupling constant (taken positive for
attractive interaction) is given as $g =-2 \hbar^2 a_{\rm{s}}/[m
\ell^2 (1-1.033 a_{\rm{s}}/\ell)]$ in the highly elongated tube
limit, with $a_s$ the two-body s-wave scattering length and $\ell$
the oscillator length of the transverse confinement in a
tube~\cite{Olshanii98}. In the tube array of lattice spacing $d$,
$\ell \sim (V_0/E_R)^{-1/4}d/\pi$. The ST (PT) of strength $t_1$
($t_2$) takes place between nearest-neighbor tubes $\left\langle
{{\bf{rr'}}} \right\rangle$.

Applying the BdG mean-field theory~\cite{DeGennes66} (which has
successfully described tube lattices without PT~\cite{Sun12} and a
variety of tube
confinements~\cite{Mizushima05,Parish07,Liu07,Sun11,Baksmaty11,Jiang11}),
we construct a mean-field Hamiltonian $H_{\rm{M}}$ by
correspondingly replacing the quartic operators in
Eq.~(\ref{eqn:HAM}) with quadratic ones coupled to three different
mean fields,
\begin{eqnarray}
{H_{{\rm{M}}}} &=& \int_z \sum\limits_{\bf{r}} \big[
\sum\limits_\sigma  {\hat \psi _{\sigma{\bf{r}}} ^\dag \big(
H_\sigma ^0 + {U_{\sigma{\bf{r}}} }\big){{\hat \psi
}_{\sigma{\bf{r}}} }} \nonumber\\
&+& \big(\Delta_{\bf{r}} \hat \psi _ {\uparrow{\bf{r}}} ^\dag \hat
\psi _ {\downarrow{\bf{r}}} ^\dag + {\rm{H}}{\rm{.c}}{\rm{.}\big)}
\big]  + \int_z {\sum\limits_{\left\langle {{\bf{rr'}}}
\right\rangle ,\sigma } {{\mathcal{T}_{\sigma{\bf{rr'}} }}} {\hat
\psi _{\sigma {\bf{r'}}}^\dag {{\hat \psi }_{\sigma {\bf{r}}}}}
}.\nonumber\\ \label{eqn:HAM_MF}
\end{eqnarray}
Here the Hartree field $U_{\sigma\bf{r}}(z)$ and the BCS gap field
$\Delta_{\bf{r}}(z)$ are standard variational fields in previous
BdG studies. We introduce a tunneling field
${{\mathcal{T}_{\sigma{\bf{rr'}} }}(z)}$ as a new ingredient to
describe the effective tunneling under the influence of both $t_1$
and $t_2$. We rotate ${H_{{\rm{M}}}}$ into the quasiparticle basis
$\hat \gamma_n$ through a Bogoliubov transformation $ {{\hat \psi
}_{\sigma\bf{r}} }({z}) = \sum_n {[{u_{n\sigma{\bf{r}}
}}({z}){{\hat \gamma }_{n\sigma }} - \sigma v_{n\sigma{\bf{r}}
}^*({z})\hat \gamma _{n, \bar \sigma }^\dag ]}$ (where $\bar
\sigma=-\sigma$) and derive extended BdG equations for the
quasiparticle wave functions $u_{n\sigma}$ and $v_{n\sigma}$ as
well as the corresponding energies $\epsilon_{n\sigma}$. The
condition $\delta \left\langle {H - TS} \right\rangle  = 0$, which
guarantees solutions of an equilibrium state at temperature $T$,
leads to the self-consistent relations
\begin{eqnarray}
{U_{\sigma{\bf{r}} }} &=& - g\left\langle {\hat\psi _{\bar
\sigma{\bf{r}} }^\dag {\hat\psi _{\bar \sigma{\bf{r}} }}}
\right\rangle  = - g\sum\limits_n {{{\left| {{u_{n\bar
\sigma{\bf{r}} }}}
\right|}^2}{f_{n \bar \sigma }}},\label{eqn:HAR}\\
{\Delta _{\bf{r}}} &=& - g\left\langle {{\hat\psi
_{{\downarrow\bf{r}}
 }}{\hat\psi _{\uparrow{\bf{r}}  }}} \right\rangle  -
{t_2}\sideset{}{'}\sum\limits_{{\bf{r'}}} {\left\langle {{\hat\psi
_{\downarrow{\bf{r'}}
 }}{\hat\psi _{ \uparrow{\bf{r'}} }}} \right\rangle
}\nonumber\\
&=& \sum\limits_n {\bigg( {-g{u_{n \uparrow {\bf{r}}}}v_{n
\downarrow {\bf{r}} }^* -
{t_2}\sideset{}{'}\sum\limits_{{\bf{r'}}} {{u_{n \uparrow
{\bf{r'}} }}v_{ \downarrow n{\bf{r'}} }^*} } \bigg){f_{n \uparrow
}}},\label{eqn:GAP}\nonumber\\\\
{\mathcal{T}_{{\bf{rr'}}\sigma }} &=&  - {t_1} - {t_2}\left\langle
{\hat\psi _{\bar \sigma {\bf{r'}}}^\dag {\hat\psi _{\bar \sigma
{\bf{r}} }}}
\right\rangle  \nonumber\\
&=&  - {t_1} - {t_2}\sum\limits_n {u_{n\bar \sigma
{\bf{r'}}}^*{u_{n\bar \sigma {\bf{r}}}}{f_{n \bar\sigma
}}},\label{eqn:TUN}
\end{eqnarray}
where ${f_{n\sigma}} = {[\exp (\epsilon_{ n\sigma}/k_B T) + 1]^{ -
1}}$ is the Fermi distribution function and
$\sum\nolimits'_{\bf{r'}}$ runs over all tubes at $\bf{r'}$ next
to $\bf{r}$. Equation (\ref{eqn:GAP}) shows that the magnitude of
the pairing gap is enhanced by $t_2$ in uniform lattices where
$\langle {{\hat\psi _{{\downarrow\bf{r}} }}{\hat\psi
_{\uparrow{\bf{r}}  }}} \rangle=\langle {{\hat\psi
_{{\downarrow\bf{r'}} }}{\hat\psi _{\uparrow{\bf{r'}} }}} \rangle$
(and would also be in trapped systems, as expected through a local
density approximation argument). This enhancement tends to
stabilize the fully paired phase against being invaded by unpaired
majority atoms; in analogy to the Meissner
effect~\cite{Meissner33}, which prevents the superconducting bulk
from being penetrated by the magnetic field. When $t_2=0$,
$\mathcal{T}=-t_1$ turns Eq.~(\ref{eqn:HAM_MF}) back to that for
the Hamiltonian with only ST (discussed in Ref.~\cite{Sun12}). We
numerically solve the BdG equation and apply the solutions to
calculate the spatial profiles of pairing gap $\Delta$, total
density $\rho=\rho_{\uparrow}+\rho_{\downarrow}$, and spin
imbalance (or magnetization)
$s=\rho_{\uparrow}-\rho_{\downarrow}$, where
$\rho_{\sigma}=\langle {\hat\psi _ \sigma ^\dag \hat\psi _ \sigma}
\rangle$ is the density profile of $\sigma$ species. In
Sec.~\ref{sec:result} we present the results for a spherically
trapped system ($\omega_r=\omega$).

\begin{figure*}[t]
\centering
\includegraphics[width=14cm]{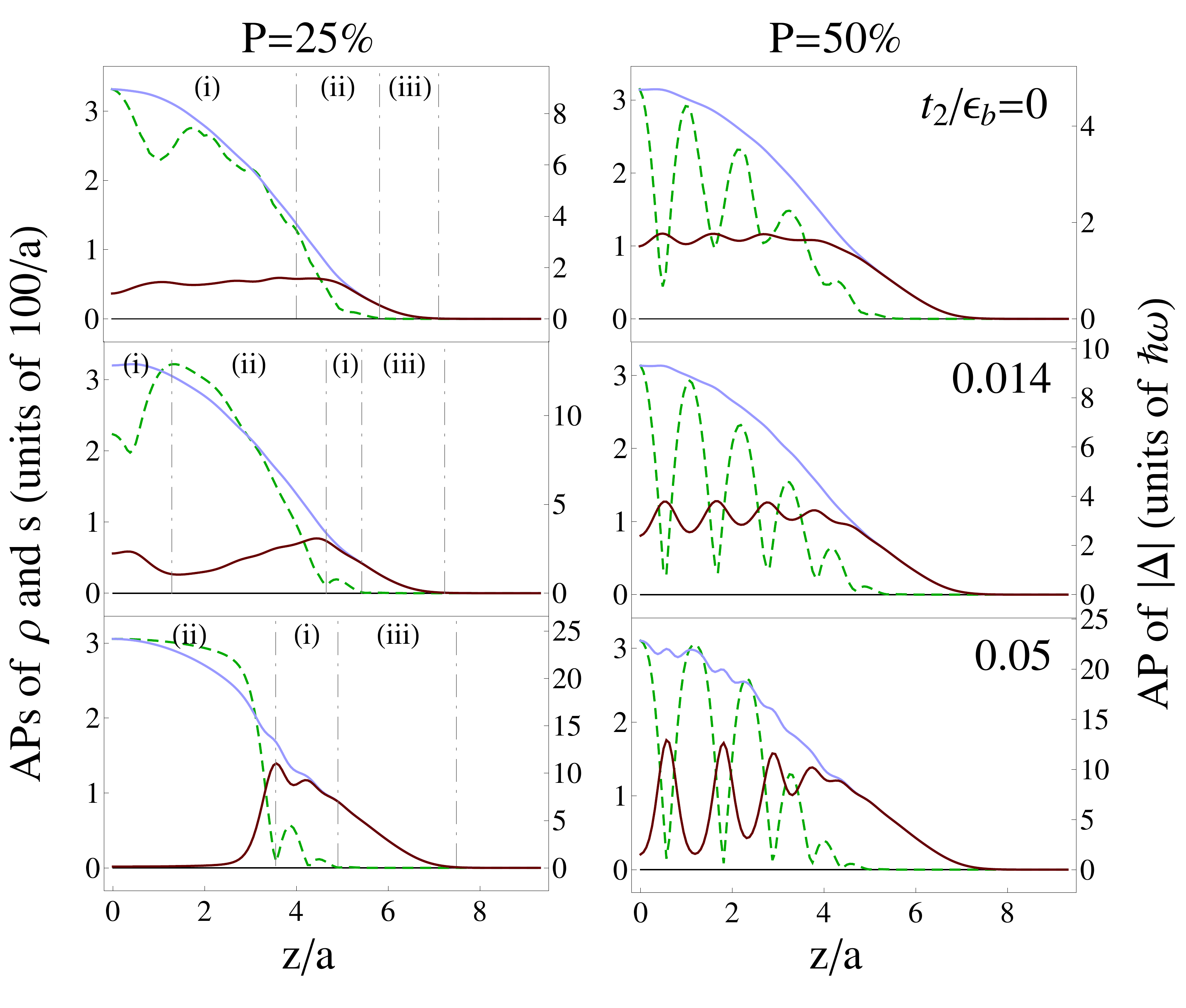}
  \caption{(Color online) Axial profiles (APs) of total density $\rho$
  and spin imbalance $s$
(solid light blue and dark red curves, respectively, axis on the
left of graph) and average magnitude of the pairing gap $|\Delta|$
(dashed green curve, axis on the right of graph) for various pair
tunneling strengths $t_2$, and global polarizations $P$. Rows from
top to bottom correspond to $t_2=0$, $0.014\epsilon_b$, and
$0.05\epsilon_b$, respectively, while the left and right columns
correspond to $P=0.25$ and $0.5$, respectively. On the left column
the dash-dotted lines demarcate regions of (i) FFLO, (ii) BCS-like
fully paired, and (iii) fully polarized states. The data obtained
were for systems of $2400$ particles in a $10\times10$ tube array
with global trapping frequency $\omega=0.0625\epsilon_b/\hbar$
(which defines the oscillator length in the $\hat z$ direction,
$\mathrm{a}=\sqrt{\hbar/m\omega}$), single-particle tunneling
$t_1=0.014\epsilon_b$, and temperature $T=0.1 \epsilon_b$. These
parameters are similar to those used in experiments.}
        \label{fig:f03}
\end{figure*}

\begin{figure}[t]
\centering
\includegraphics[width=8.7cm]{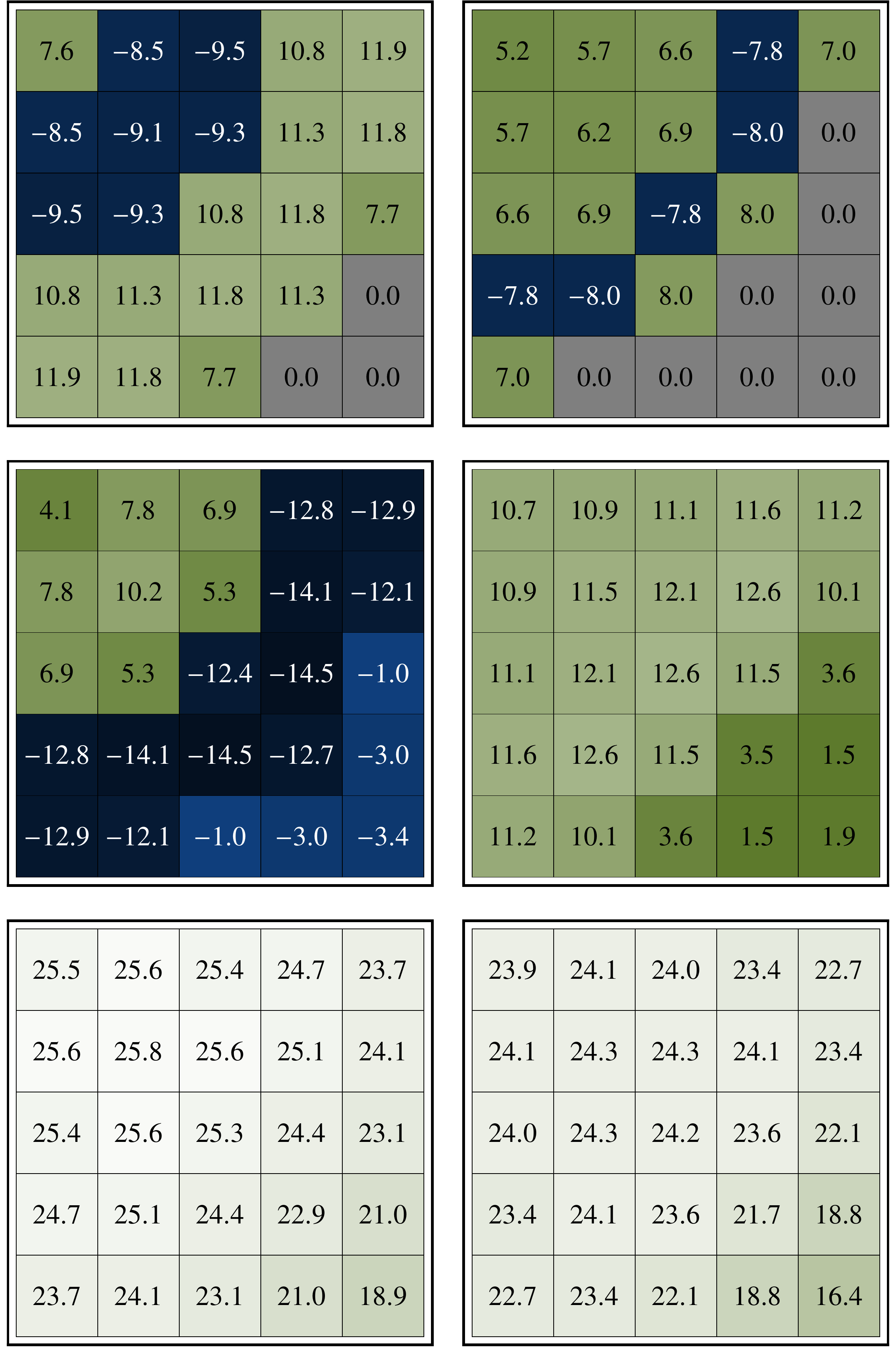}
  \caption{(Color online) Value of the gap function
  (in units of $\hbar \omega$) at the center of each tube ($z=0$) for various $t_2$ and $P$
(convention as presented in Fig.~\ref{fig:f03}). Here we show data
for $5 \times 5$ tubes in the fourth quadrant of the $10 \times
10$ tube array, in which the top left entry of each panel
corresponds to the most central tube. The other quadrants are
similar due to fourfold rotational symmetry.}
        \label{fig:f04}
\end{figure}

\section{Results}\label{sec:result}
From now on, we take a realistic setup $d=-a_s=0.5 \mu$m for
$^{6}$Li systems in the Feshbach resonant regime and use the
binding energy $\epsilon_b=mg^2/4\hbar^2$ as the energy unit for
the following results. We look at the influence of $t_2$ at fixed
$t_1=0.014 \epsilon_b$, the latter corresponding to a typical
lattice depth of $7 E_R$ and thus into the dimensional crossover
regime. In Fig.~\ref{fig:f03} we plot the axial profiles of
$\rho$, $s$, and the average of $|\Delta|$ by tracing out the
${\bf{r}}$ degree of freedom. The first and second columns
correspond to a lower global polarization of $P=25\%$ (LP) and a
higher one of $P=50\%$ (HP), respectively\footnote{The global
polarization $P$ is defined as the ratio of total imbalance to the
total number of particles. The LP case we focus on herein is
somewhat higher than the critical polarization of $15\%$--$18\%$
verified in experiments.}. From top row to bottom, $t_2$ is chosen
to be either zero, comparable to $t_1$, or larger than $t_1$,
respectively. We see that in the LP case at $t_2=0$, the axial
profile exhibits (i) an FFLO center with oscillatory $\Delta$,
(ii) a BCS-like shoulder with non oscillatory $\Delta$, and (iii)
a normal tail having zero $\Delta$. At the intermediate $t_2$
value, this trilayered structure remains. However, the FFLO center
shrinks, the BCS-like region extends toward the center accompanied
by a drop in imbalance, and the normal tail grows. This indicates
a transfer of unpaired majority atoms from the center to the tail,
implying an enhancement of a Meissner-like effect in the central
region. We notice that the gap profile develops small ripples
between the BCS-like shoulder [(ii)] and the normal tail [(iii)],
suggesting the incipience of an FFLO layer [(i)] here. At the
large $t_2$ value, the FFLO center is completely conquered by the
BCS-like state and disappears, leaving a large fully polarized
tail and a thin FFLO layer in between them. Because the FFLO and
BCS centers are distinctive of one-~\cite{Liao10} and
three-dimensional~\cite{Zwierlein06,Partridge06} trapped systems,
respectively, this result shows the evolution of the system from
1D toward 3D, driven by $t_2$ (compared with increasing $t_1$).

In the HP case, the system always has a center with oscillatory
$\Delta$ and a fully polarized normal tail. In the
oscillatory-pairing region, the imbalance profile exhibits
characteristic out-of-phase oscillations, with the concurrence of
local minima (maxima) of $s$ and local maxima (minima) of
$|\Delta|$. This behavior is due to the competition between
superfluid and ferromagnetic orders. An increase in $t_2$ enhances
this competition, augmenting the magnitude of the out-of-phase
oscillations and repelling a portion of the unpaired majority to
the normal tail region. At large $t_2$ ($=0.05\epsilon_b$), the
oscillations are large enough that the minima of $s$ are almost
zero. Such a case is less like an FFLO state (oscillatory pairing
accompanied by finite polarization), but more like spatial
alternation of fully paired superfluid and highly polarized normal
gas. This phenomenon, which is analogous to the phase separation
in the LP case, is taken as a signature of the dimensional
crossover between 1D and 3D at higher polarizations. We also
notice that the structure of the profiles is reminiscent of that
of a system with vortex cores embedded in a superfluid bulk.

\begin{figure}[t]
\centering \subfigure[]{\includegraphics[width=6.5cm]{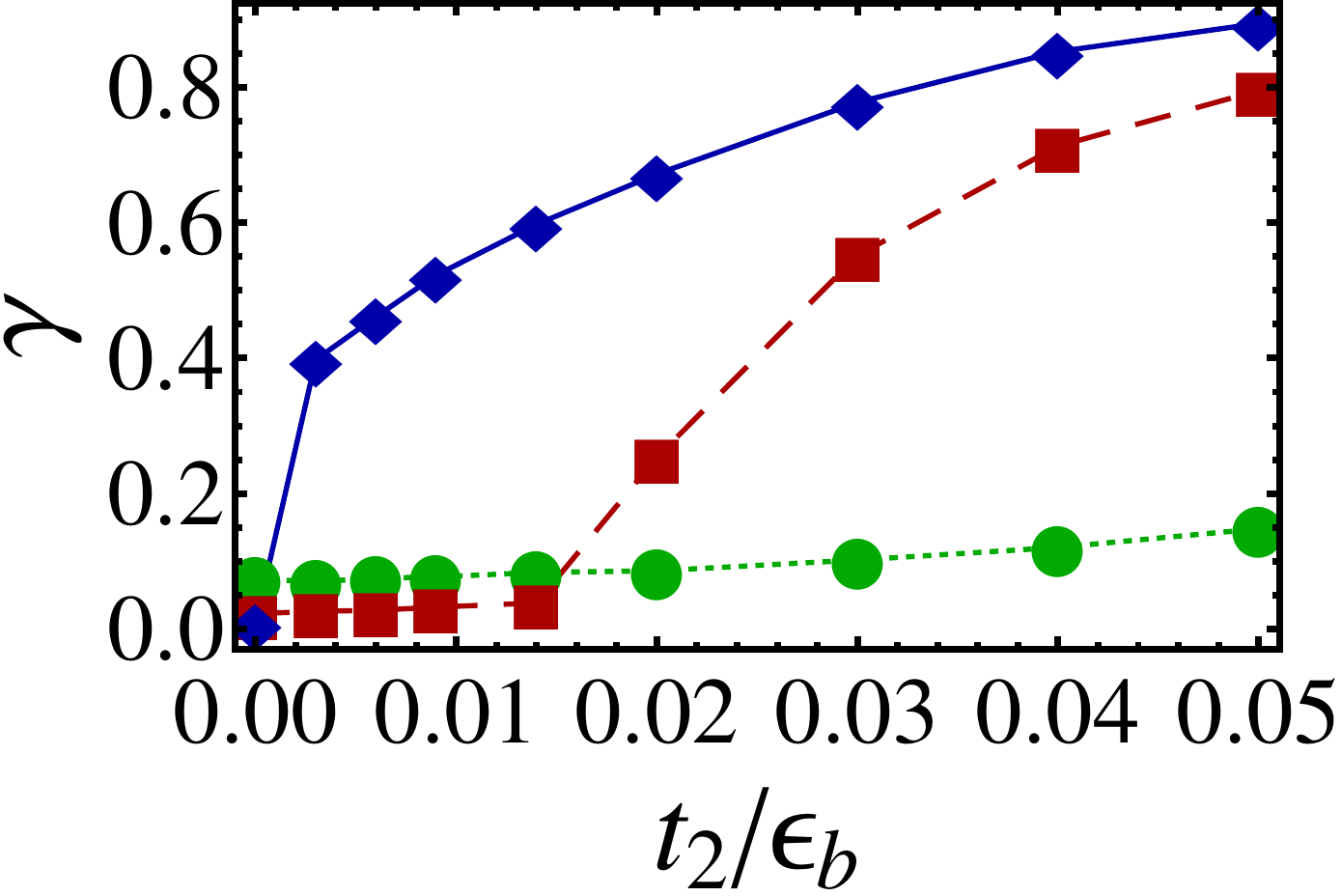}}
\subfigure[]{\includegraphics[width=6.5cm]{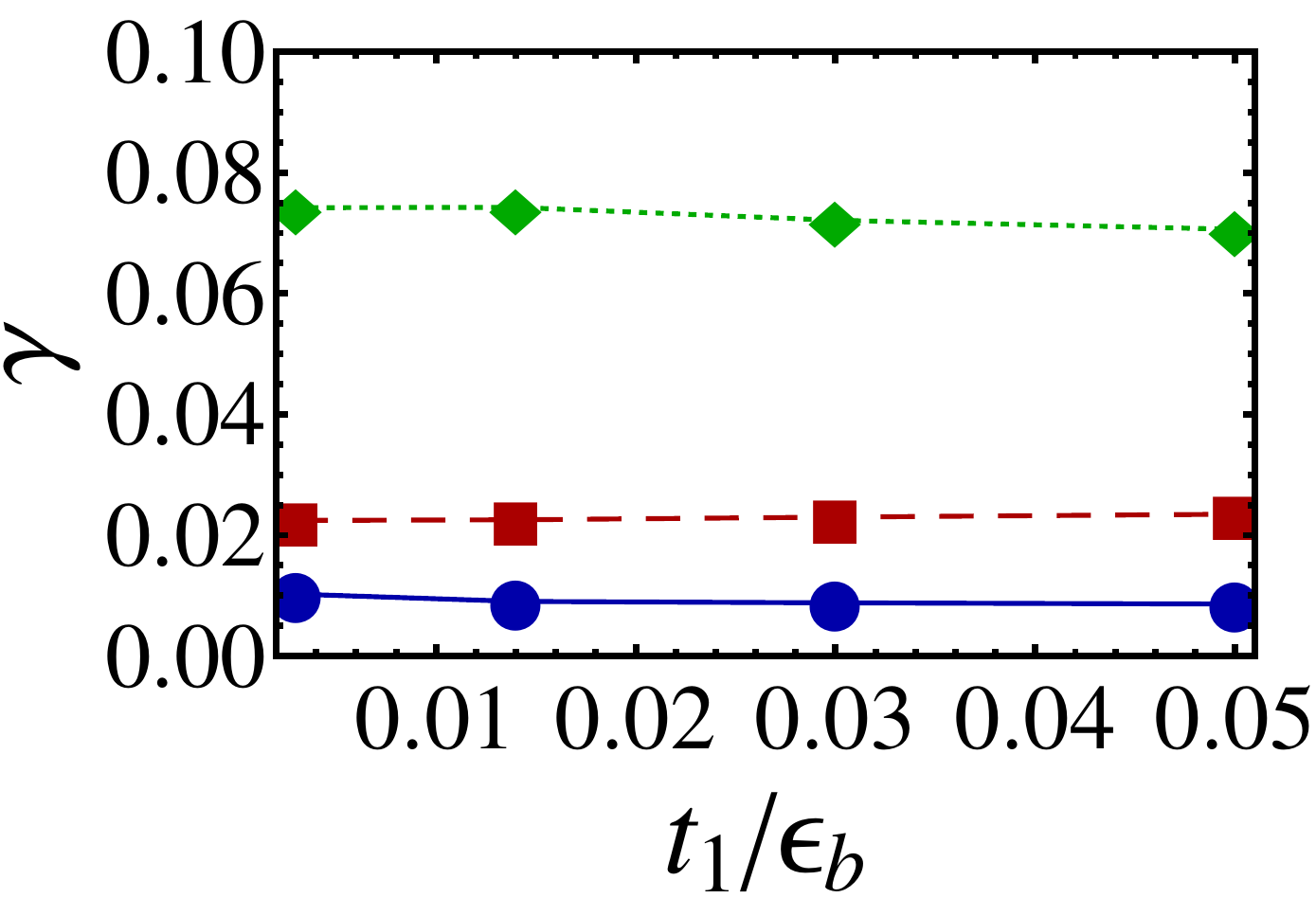}}
  \caption{(Color online) Combined fraction of particles
in the highly paired ($s/\rho<5\%$) and highly polarized
($s/\rho>95\%$) regions $\gamma$ vs (a) $t_2$ or (b) $t_1$ with
the other fixed. The solid blue, dashed red, and dotted green
curves represent cases with global polarization $P=12.5$\%,
$25$\%, and $50$\%, respectively.}
        \label{fig:f05}
\end{figure}

We find that PT affects the pairing order not only along but also
across the tubes. Figure \ref{fig:f04} shows the value of the gap
function at the center of each tube ($z=0$) in a $10\times10$ tube
array. The left (right) column corresponds to the LP (HP) case,
while, from top to bottom, rows correspond to zero, intermediate,
and large $t_2$, respectively (as in Fig.~\ref{fig:f03}). We see
that at zero $t_2$ the sign of $\Delta$ changes, indicating an
oscillatory behavior across the tubes. In the LP case when $t_2$
increases, the oscillating nodes appear in a more off-center
region, as discussed for the axial profiles along the tubes. At
large $t_2$, there is no oscillation of $\Delta$ across tubes in
both LP and HP cases, showing the suppression of FFLO order. We
notice in Figs.~\ref{fig:f03} and \ref{fig:f04} that $t_2$
enhances the maximum magnitude of the gap function, as expected
from Eq.~(\ref{eqn:GAP}). This enhancement raises the critical
temperature above which the pairing order vanishes and hence
agrees with the increase of the superfluid transition temperature
in quasi-one-dimensional systems~\cite{Zhao08}.

Finally, we look at the phase separation of fully paired and fully
polarized regions as a function of $t_2$. We consider the combined
fraction of particles in the highly paired ($s/\rho<5\%$) and
highly polarized ($s/\rho>95\%$) regions of the axial profiles;
$\gamma  \equiv \int_z {\rho [\theta (0.05 - s/\rho ) + \theta
(s/\rho  - 0.95)]} /\int_z \rho$, where $\theta$ is the step
function. The larger $\gamma$ is, the stronger phase separation
the system shows. Figure 5(a) shows that $\gamma$ monotonically
increases with $t_2$ at three various polarizations when $t_1$ is
fixed. In the cases of $P=12.5\%$ and $25\%$ the sudden changes
indicate the occurrence of the BCS-like center replacing the FFLO
center. For comparison we plot also $\gamma$ vs $t_1$ at fixed
$t_2$ in Fig.~5(b) and observe that $\gamma$ shows almost no
change at the three polarizations. This result highlights that it
is $t_2$, rather than $t_1$, that accounts for the phase
separation and hence is essential for the correct model describing
the physics at the incipience of the dimensional crossover regime.

\section{Conclusion}\label{sec:conclusion}
By considering the microscopic physics of cold atomic systems, we
have incorporated both ST and PT processes to effectively model
imbalanced fermionic superfluids in an array of one-dimensional
tubes at the incipience of the dimensional crossover. Our
calculations show that the PT strength is a main factor for the
evolution of the system profiles deviating from the
one-dimensional limit, which exhibits a central FFLO state, toward
the development of three-dimensional signatures, including a
central fully paired state in the LP case and spatial separation
between fully paired and fully polarized states in both LP and HP
cases. These features are reflected in the directly observed
density profiles and the pairing orders that can be probed in
time-of-flight
experiments~\cite{Kinoshita04,Swanson12,Lu12,Bolech12}. Our model
can be easily generalized to incorporate higher-order, higher-band
or interband processes~\cite{Duan05,Mathy09}, which are expected
to be of further help in the investigation of the system's
transition to the continuous three-dimensional limit.

We are grateful to T. Giamarchi for interesting discussions and
the Kavli Institute for Theoretical Physics where these took place
(NSF Grant No. PHY05-51164). We thank R. Hulet and his group for
valuable discussions and the sharing of preliminary experimental
data~\cite{Hulet12}. This work was supported by the DARPA-ARO
Award No. W911NF-07-1-0464.

\end{document}